\documentclass{emulateapj}
 
\usepackage{graphicx}
\usepackage{amssymb}
\usepackage{epstopdf}
\usepackage{amsmath}

\newcommand{\sn}{\mbox{SN\,2011dh}}
\newcommand{\xray}{\mbox{X-ray}}
\newcommand{\swift}{\mbox{\it Swift}}
\def\nh{\mbox{$N_{\rm H}$}}

\newcommand{\chandra}{\mbox{Chandra}}

\newcommand{\erg}{\mbox{erg}}

\newcommand{\persec}{\mbox{s$^{-1}$}}
\newcommand{\percmsq}{${\rm cm}^{-2}$}
\newcommand{\cgsflux}{\erg\,\percmsq\,\persec}
\newcommand{\perct}{\mbox{ct$^{-1}$}}

\newcommand{\perksec}{\mbox{ks$^{-1}$}}

\begin{document}

\title{An Early  \&\ Comprehensive Millimeter and Centimeter Wave and X-ray Study of
  Supernova 2011dh: A Non-Equipartition Blastwave Expanding into A Massive Stellar Wind}


\author{
Assaf Horesh\altaffilmark{1},
Christopher Stockdale\altaffilmark{2,3},
Derek B. Fox\altaffilmark{4}, 
Dale A. Frail\altaffilmark{5}
John Carpenter\altaffilmark{1}, 
S.~R.~Kulkarni\altaffilmark{1},
Eran O. Ofek\altaffilmark{1,6},
Avishay Gal-Yam\altaffilmark{6},
Mansi M. Kasliwal\altaffilmark{1,7}, 
Iair Arcavi\altaffilmark{6},
Robert Quimby\altaffilmark{8},
S. Bradley Cenko\altaffilmark{9},
Peter E. Nugent\altaffilmark{10,9},
Joshua S. Bloom\altaffilmark{9,10},
Nicholas M. Law\altaffilmark{11},
Dovi Poznanski\altaffilmark{12},
Evgeny Gorbikov\altaffilmark{12},
David Polishook\altaffilmark{6,13},
Ofer Yaron\altaffilmark{6},
Stuart Ryder\altaffilmark{14},
Kurt W. Weiler\altaffilmark{15},
Franz Bauer\altaffilmark{16,17},
Schuyler D. Van Dyk\altaffilmark{18},
Stefan Immler\altaffilmark{19,20,21},
Nino Panagia\altaffilmark{22,23,24},
Dave Pooley\altaffilmark{25},
Namir Kassim\altaffilmark{26}
}

\altaffiltext{1}{Cahill Center for Astrophysics, California Institute
  of Technology, Pasadena, CA, 91125, USA}
\altaffiltext{2}{Department of Physics, Marquette University, PO Box 1881, Milwaukee, WI
53201.}
\altaffiltext{3}{Homer L. Dodge Department of Physics \& Astronomy, The
University of Oklahoma, Norman, OK 73019.}
\altaffiltext{4}{Astronomy and Astrophysics, Eberly College of
  Science, The Pennsylvania State University, University Park, PA
  16802, USA}
\altaffiltext{5}{National Radio Astronomy Observatory, P.O. Box 0, Socorro, NM 87801, USA}
\altaffiltext{6}{Benoziyo Center for Astrophysics, Faculty of Physics,
  The Weizmann Institute of Science, Rehovot 76100, Israel}
\altaffiltext{7}{Carnegie Institution for Science, 813 Santa Barbara St, Pasadena, CA, 91101, USA}
\altaffiltext{8}{IPMU, University of Tokyo, Kashiwanoha 5-1-5, Kashiwa-shi, Chiba, Japan}
\altaffiltext{9}{Department of Astronomy, University of California, Berkeley, CA 94720-3411, USA}
\altaffiltext{10}{Computational Cosmology Center, Lawrence Berkeley National Laboratory, 1 Cyclotron Road, Berkeley, CA 94720, USA}
\altaffiltext{11}{Dunlap Institute for Astronomy and Astrophysics,
  University of Toronto, 50 St. George Street, Toronto M5S 3H4,
  Ontario, Canada}
\altaffiltext{12}{School of Physics and Astronomy, Tel-Aviv University, Tel-Aviv 69978, Israel}
\altaffiltext{13}{Department of Earth, Atmospheric, and Planetary Sciences, Massachusetts Institute of Technology, Cambridge, MA 02139, USA.}
\altaffiltext{14}{Australian Astronomical Observatory, P.O. Box 915,
  North Ryde, NSW 1670, Australia}
\altaffiltext{15}{Computational Physics Inc., 8001 Braddock Road,
  Suite 210, Springfield, VA 22151-2110, USA}
\altaffiltext{16}{Pontificia Universidad Cat\'{o}lica de Chile, Departamento de Astronom\'{\i}a y Astrof\'{\i}sica, Casilla 306, Santiago 22, Chile}
\altaffiltext{17}{Space Science Institute, 4750 Walnut Street, Suite 205, Boulder, Colorado 80301}
\altaffiltext{18}{Spitzer Science Center/Caltech, Mailcode 220-6,
  Pasadena, CA 91125, USA}
\altaffiltext{19}{Astrophysics Science Division, NASA Goddard Space Flight Center, Greenbelt, MD 20771}
\altaffiltext{20}{Department of Astronomy, University of Maryland, College Park, MD 20742}
\altaffiltext{21}{Center for Research and Exploration in Space Science
  and Technology, NASA Goddard Space Flight Center, Greenbelt, MD
  20771}
\altaffiltext{22}{Space Telescope Science Institute, 3700 San Martin
  Drive, Baltimore, MD 21218, USA}
\altaffiltext{23}{INAF–CT, Osservatorio Astrofisico di Catania, Via S. Sofia 78, I-95123 Catania, Italy}
\altaffiltext{24}{Supernova Ltd, OYV $\#131$, Northsound Rd., Virgin Gorda, British Virgin Islands}
\altaffiltext{25}{Department of Astronomy, University of Texas, Austin, TX 78712, USA}
\altaffiltext{26}{Naval Research Laboratory, Code 7213, Washington, DC
  20375-5320, USA}

\begin{abstract}

Only a handful of supernovae (SNe) have been studied in
multi-wavelength from radio to X-rays, starting a few days after
explosion. The early detection and classification of the nearby type
IIb SN\,2011dh/PTF\,11eon in M51 provides a unique opportunity to
conduct such observations. We present detailed data obtained at the youngest phase
ever of a core-collapse supernova (days 3 to 12 after explosion) in
the radio, millimeter and X-rays; when combined with
optical data, this allows us to explore the early evolution of the SN
blast wave and its surroundings. Our analysis shows that the expanding
supernova shockwave does not exhibit equipartition
($\epsilon_{\rm e}/\epsilon_{\rm B}\sim 1000$), and is expanding
into circumstellar material that is consistent with a density
profile falling like $R^{-2}$. Within modeling uncertainties we find
an average velocity of the fast parts of the ejecta of 15,000$\pm$1800 km/s, contrary to previous
analysis. This velocity places SN 2011dh in an intermediate blast-wave regime
between the previously defined compact and extended SN IIb
subtypes. Our results highlight the importance of early ($\sim1$ day)
high-frequency observations of future events. Moreover, we show the importance
of combined radio/X-ray observations for determining the microphysics ratio $\epsilon_{\rm e}/\epsilon_{\rm B}$.

\end{abstract}

\section{Introduction}

SN\,2011dh  in the nearby galaxy Messier 51 was discovered on UT
2011 May 31.893 by A. Riou; detected on June 01.19 UT by the Palomar
Transient Factory (PTF, Law et al. 2009, Rau et al. 2009); and
rapidly spectroscopically classified as a type IIb supernova (Arcavi
et al. 2011a; 2011b).

The proximity of M51 (as in Arcavi et al. 2011b we assume a distance,
$d=8.03\pm 0.77$ Mpc) motivated searches for the progenitor star.
A putative massive star detected by the Hubble Space Telescope in pre-explosion images of M51 very close to the SN position
was
described by Maund et al. (2011) and Van Dyk et al. (2011).  The
interpretation of the progenitor is uncertain. Based on the rapid cooling of the expanding SN
ejecta, Arcavi et al. (2011b) argue that the progenitor star must
have a radius smaller than that of typical red supergiants or the
supergiant progenitor of the well-studied Type IIb SN 1993J, see e.g.,
Weiler et al. (2007)
(R$<10^{13}$ cm).  In contrast, the astrometrically coincident star
is  a F8 supergiant. Reconciling these constraints means that
the supergiant is either unrelated to the SN or was a companion of the
(now erstwhile) progenitor star.

For core collapse supernovae the interaction of the blast wave with the
circumstellar medium can generate detectable radio and X-ray emission.
Additionally the fastest moving ejecta is related to the size of the 
progenitor (the more compact, the higher the velocity; Chevalier \&\ Soderberg
2010). These two
properties motivate early radio and X-ray observations. Thus, we began a program
of centimeter and millimeter wave observations of SN\,2011dh with
EVLA and CARMA respectively. We initiated X-ray observations with the
{\it Swift} Observatory. Following the end of our early monitoring a
long-term program at EVLA was launched by A. Soderberg and
collaborators. Their results are reported in a recent paper by Krauss et
al. (2012). 

Soderberg et al. (2012) analyzed the {\it Swift} X-ray and  two epochs
(day 4 and day 17)
of radio observations of SN\,2011dh, and found an expansion speed of
$\approx 0.1c$ and a mass-loss rate of $6\times 10^{-5}\,M_\odot\,{\rm
yr}^{-1}$.  With these inferences they concluded that SN\,2011dh is probably a type cIIb SN, namely a SN that originated
from a compact progenitor star as opposed to one that originated
from an extended progenitor star (eIIb); see Chevalier \&\ Soderberg
(2010) and references therein for explanation of these proposed
sub-types. A similar result was found by Krauss et al. (2012).

Here we present a full set of centimeter and millimeter wave
observations (from day 3 to day 12 after explosion) 
as well as an analysis of the Swift and
Chandra X-ray observations of SN\,2011dh. To our knowledge, these
represent the most comprehensive (essentially daily) set of pan radio (5 GHz to 100 GHz)
observations obtained at early times of a core collapse supernova. As a result we
are able to probe the circumstellar matter at smaller radii (wherein
one can expect to find deviations from homologous flows) and in
principle directly infer the density of circumstellar matter. 

The paper is organized as follows. 
The millimeter (CARMA) and centimeter (EVLA) observations are summarized in
\S\ref{sec:CARMA-EVLA}. In \S\ref{sec:SSA} we carry out a standard analysis for the radio
observations assuming a synchrotron self-absorbed model. We then summarize the
X-ray observations (\S\ref{sec:Xray}) and present a combined radio+X-ray
analysis (which incorporates inverse Compton scattering of optical photons
to the X-ray band) in \S\ref{sec:Combined}. We present our conclusions
as well as review the returns obtained from early millimeter wave observations
of core collapse supernovae (\S\ref{sec:Conclusions}). Such a review is
timely given the imminent operation of the Atacama Large Millimeter Array (ALMA).

\section{CARMA and EVLA observations}
\label{sec:CARMA-EVLA}

\begin{figure*}[!ht] 
\centering
\includegraphics[width=0.95\textwidth]{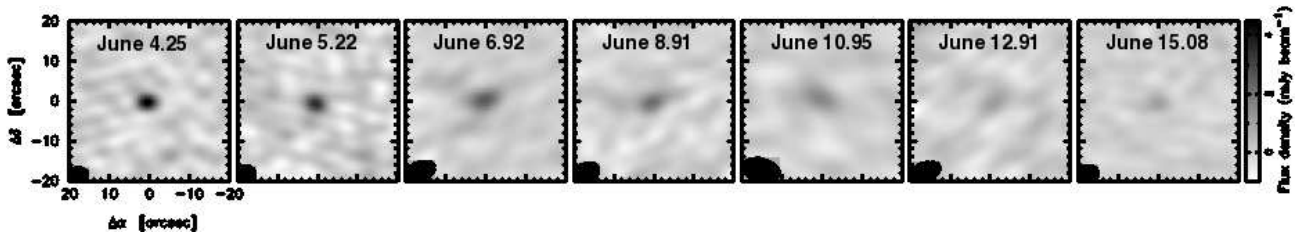}
\caption{
\small 
CARMA image cutouts  of SN\,2011dh in the 3-mm band. The source at the
  center is SN\,2011dh. The image cutouts are 40'' on a side
centered around SN\,2011dh's most precise position, based on our EVLA
observations, at Right Ascension=13:30:05.104 and Declination
=47:10:10.915 (J2000.0). The positional error is +/-0.01 arcsec in
each axis (see text).
}
\label{fig:CARMAMontage}
\end{figure*}

Starting June 4, 2011, observations were
undertaken with CARMA, centered on either 107 GHz or 93 GHz with an 8 GHz
bandwidth in the D array configuration. 
For flux calibration we used either Uranus (whose flux was
bootstrapped from observations of 3C345) or Mars. The compact source
J1153+495 was used as a phase calibrator. The CARMA data were reduced using the MIRIAD reduction
software\footnote{http://bima.astro.umd.edu/miriad/}. 
The log of observations can be found in Table~\ref{tab:RadioLog} and a
montage of CARMA detections is shown in Figure~\ref{fig:CARMAMontage}.

Observations with the EVLA were undertaken in the framework of 
our program  ``PTF Transients in the
Local Universe'' (PI: Kasliwal) as well as a ToO program to observe Type II Supernovae (PI:  C. 
Stockdale). The array configuration was A. 3C286 served as the flux
calibrator. We used the AIPS software to reduce the EVLA data. 
The first EVLA detection of the SN was in both the K (22.46 GHz) and Q
(43.46 GHz) bands at fluxes of 2.6 mJy and 6.95 mJy, respectively. 
The log of the observations can be found in Table~\ref{tab:RadioLog}.
 \begin{table*}[!ht]

\caption{Summary of early radio observations of SN\,2011dh}
\smallskip
\begin{center}
\begin{tabular}{lrl}
\hline
\noalign{\smallskip}
Day & Frequency   & Flux    \\
               &  [GHz] & [mJy] \\
\noalign{\smallskip}
\hline
\noalign{\smallskip}

4.21 &	 4.8 &  $\leq0.17$ \\
4.21	 &    7.4	& 0.203  $\pm$ 0.036 $\pm$ 0.01\\
4.25 &	 22.5& 	 2.6	$\pm$ 0.07 $\pm$ 0.26\\
4.23 &	 43.2 &	 6.95  $\pm$  0.17 $\pm$ 0.7\\ 
4.08 &	 107	 &4.55  $\pm$  0.33 $\pm$ 0.46\\
4.08 &	 230	 & $\leq 3$       \\
5.01 &	 8.5  &  0.455  $\pm$ 0.046 $\pm$ 0.023\\ 
5.12 &	 22.5 &	 3.95   $\pm$ 0.07  $\pm$ 0.4\\
5.10 &	 43.2 &	 6.47  $\pm$  0.14  $\pm$ 0.65\\
5.22 &	 107	& 3.66  $\pm$  0.35 $\pm$ 0.36 \\
6.92	 &93	& 2.52 $\pm$	 0.27 $\pm$ 0.25\\
7.12	 &8.5	& 1.06 $\pm$	 0.03 $\pm$ 0.053\\
7.20	 &43.2	& 6.42 $\pm$	 0.17 $\pm$ 0.64 \\
7.22	 &22.5	& 6.89 $\pm$	 0.06 $\pm$ 0.69\\
8.91	 &93	& 1.84 $\pm$ 0.31 $\pm$ 0.18\\
9.02   &  22.5 &  7.46 $\pm$    0.04  $\pm$ 0.75\\
9.02     &8.5 &   1.58    $\pm$  0.03 $\pm$ 0.079\\
9.06     &5.0  &  0.42   $\pm$ 0.03 $\pm$ 0.021\\
9.12	 &33.6	& 7.49 $\pm$	 0.06 $\pm$ 0.75\\
10.95 &   93     & 1.61 $\pm$  0.30 $\pm$ 0.16\\
11.01   & 43.2 &  3.19 $\pm$  0.15 $\pm$ 0.32\\
11.03   & 22.5  & 8.17   $\pm$  0.05 $\pm$ 0.82\\
11.98  &  8.5   & 3.15   $\pm$ 0.06 $\pm$ 0.158\\
12.02  &  5.0   & 1.22 $\pm$   0.03 $\pm$ 0.061\\
12.91  &  93     & 1.06  $\pm$  0.31 $\pm$ 0.106\\
15.08  &  93  &  1.05  $\pm$  0.17 $\pm$ 0.105 \\

\noalign{\smallskip}
\hline
\smallskip
\end{tabular} 
\label{tab:RadioLog}
\end{center}
{\small
Notes - Day is given in UT days in June
2011. The errors presented in the table represent the rms error from
each image and a systematic calibration error ($5\%$ for $\nu<20$ GHz
and $10\%$ for $\nu>20$ GHz) that should be combined in quadrature.}     
\end{table*}

\section{Standard Synchrotron Analysis}
\label{sec:SSA}

The radio emission arises from relativistic
particles, which are accelerated at the shock, gyrating in magnetic
fields, generated in the
post-shock gas (Chevlaier 1982; Chevalier 1998; Weiler et al. 2002). A
large body of work confirms that (for most supernovae) the early radio
emission can be described by a synchrotron self-absorption (SSA) model
(Chevalier 1998). While internal free-free absorption might also be present at
early times, the lack of a steeper spectrum at the optically thick
part of the spectrum suggests that SSA is dominant (see also Appendix 1). Hence, throughout
this paper we adopt Chevalier (1998) SSA model.

Assuming a simple power law for the energy of the relativistic
particles, the synchrotron emission for a self-absorbed source can be described by
\begin{equation}
S_{\nu}\propto \frac{\pi
  R^{2}}{D^{2}}B^{-1/2}\nu^{5/2}
\label{eq:SSA1}
\end{equation}
for frequency, $\nu$ below $\nu_a$, where $\nu_a$ is the
frequency at which the optical depth from synchrotron self-absorption is unity, and 
\begin{equation}
S_{\nu}\propto \frac{4 \pi
  f R^{3}}{3 D^{2}}N_{0}B^{(p + 1)/2}\nu^{-(p - 1)/2},
\label{eq:SSA2}
\end{equation} 
for frequencies above $\nu_{a}$.
Here, $B$ is the the strength of the magnetic field, $R$ is
the radius of the blast-wave, $D$ is the distance from the observer to
the supernova, $f$ is the volume fraction of the radio emitting
region, and the energy spectrum of the relativistic particles
is given by a power law, 
$N(E)=N_{0}E^{-p}$. 

The SSA model does not attempt to predict the absolute or even
the relative 
fraction of the two key components, the energy in relativistic electrons
and the strength of the magnetic field.  
However, the minimum total energy will be
achieved at equipartition, i.e., when the energy in the relativistic electrons is equal to that of the
magnetic fields or $f_{\rm eB}=1$ (see Readhead 1994). Here $f_{\rm eB}$ is the ratio
of the energy in relativistic electrons to that of the magnetic fields.

The measurement of the single-epoch SSA spectrum (both the optically
thick and the optically thin parts) can be inverted to yield the
radius and the magnetic field
at that epoch (Chevalier 1998). For $p=3$ (which is the relevant case
here; see \S 3.1): 
\begin{eqnarray}
R_{p}&=&8.8 \times 10^{15} f_{\rm
  eB}^{-1/19}\left(\frac{f}{0.5}\right)^{-1/19}\left(\frac{S_{p}}{\rm Jy}\right)^{9/19}
\left(\frac{D}{\rm Mpc}\right)^{18/19} \nonumber\\
& & \left(\frac{\nu_{p}}{\rm 5\,GHz}\right) ^{-1}{\rm cm}
\label{eq:R}
\end{eqnarray}
and
\begin{eqnarray}
B_{p}&=&0.58
f_{\rm
  eB}^{-4/19}\left(\frac{f}{0.5}\right)^{-4/19}\left(\frac{S_{p}}{\rm Jy}\right)^{-2/19}
\left(\frac{D}{\rm Mpc}\right)^{-4/19}\nonumber\\
& &\left(\frac{\nu_{p}}{\rm 5\,GHz}\right)  {\rm
  G}.
\label{eq:B}
\end{eqnarray}
Here, $S_{p}$ is the peak flux\footnote{The peak flux marks the
  transition of the spectrum from optically thick to optically thin.}, and $\nu_{p}$ is the peak
frequency. We assume $f=0.5$ (see Chevalier \& Fransson 2006).

Equations~\ref{eq:SSA1} and \ref{eq:SSA2} are adequate to describe the
broad-band spectrum at any given epoch, as long as the emission is
SSA. The dynamics of the supernova
shell (which in turn depend on the velocity profile of the blast wave
and the radial density distribution of the circumstellar medium)
determine $R$ and $\nu_a$. These dependencies are generalized by
allowing
for power-law variations in key quantities (Weiler et al. 2002) and
leading to the following equations: 
\begin{equation}
S=K1 \left(\frac{\nu}{{\rm 5\,GHz}}\right)^{\alpha}
\left(\frac{t-t_{0}}
{\rm
  1\,day}\right)^{\beta}\left(\frac{1-e^{-\tau_{SSA}}}{\tau_{SSA}}\right), 
\label{eq:S}
\end{equation}
where the absorption expression  describes an internal absorption by material
mixed with the emitting component and assumes planar geometry. The
SSA optical depth is described by
\begin{equation}
\tau_{SSA}=K5 \left(\frac{\nu}{\rm 5\,GHz}\right)^{\alpha-2.5}
\left(\frac{t-t_{0}}
{\rm 1\,day}\right)^{\delta ^{''}},
\label{eq:tau}
\end{equation}
where both $K1$ and $K5$ are proportionality constants that can be
determined by fitting the data, and $\delta^{''}$ describes the time
dependence of the optical depth.

\subsection{Blast-wave physical parameters at single epochs}
\label{sec:equip}
We analyze observations from the first five epochs: 2011 June 4, 5, 7,
9, and 11. During the first four epochs, data were obtained within a
span of $\approx 4$ hours and thus are considered to be nearly
simultaneous such that we assigned a central average time to all flux
densities of a given epoch. We note that the last C-, and X-band
observation were made on June 12 rather than on June 11 and therefore
we obtain the C-, and X-band fluxes for June 11 by
interpolation. For each epoch we fit the EVLA-CARMA spectrum to the
SSA model characterized by $p$, $S_p$ and
$\nu_p$ (Equations \ref{eq:SSA1} and \ref{eq:SSA2}). We assume
equipartition. 
For each epoch, the SSA model is
well fitted by the data. The electron energy index, 
$p$, averaged over the five epochs we analyze is $p=3.00\pm 0.18$.  
Next, substituting our fitted parameters into Equations~\ref{eq:R} and \ref{eq:B} 
we derive the magnetic field ($B)$ and radius ($R$) at each of our
five epochs.

\begin{figure}
\centering
\includegraphics[width=0.5\textwidth]{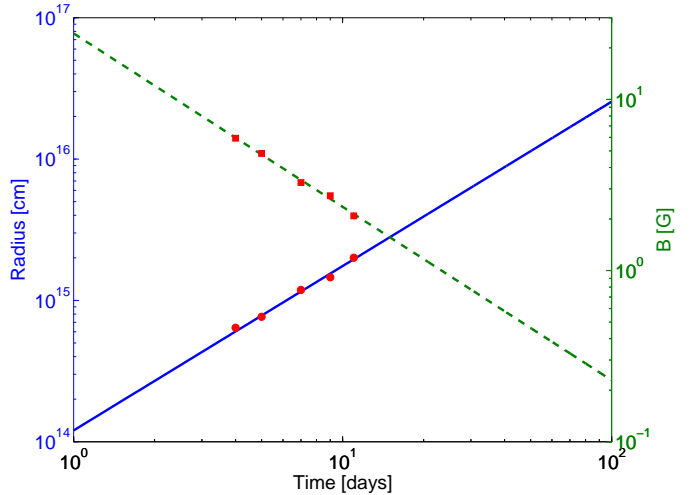}
\caption{The blast-wave radius (red circles) and magnetic field
  (red squares) estimates at 5 epochs. We assume an explosion date of May
  31.5, 2011. Relative to this explosion date the epochs are [3.7,
  4.6, 6.6, 8.5, 10.5] days. 
The corresponding values are: 
${\rm R}\approx[0.64, 0.77,
1.19, 1.46, 2.01]\times 10^{15} {\rm cm}$ and ${\rm B}\approx[5.9, 4.8, 3.3, 2.7, 2.1]$G. 
Given the uncertainty in
the peak fluxes, peak frequencies and $p$, we estimate the radius
and magnetic field uncertainties to be $11\%$ and $9\%$,
respectively.
The curves are power law fits
  with $R\propto t^{1.14\pm 0.24}$ (solid blue line) and $B\propto
  t^{-1\pm 0.12}$ (dashed green line).}
\end{figure}

As shown in Figure 2, the
blast-wave radius, $R\propto t^m$ where $m=1.14\pm0.24$ while the magnetic field
is proportional to $t^{-1\pm 0.12}$. 
Theoretically, the value of $m$ depends on the density structure of
the blast wave. In particular, once the blast-wave decelerates,
$m=1-(1/ \eta)$ where $\eta$ is the velocity power-law index of the
blastwave ($R\propto v^{-\eta}$). $\eta$ can be described as a function
of the progenitor star polytropic index, $n$, where
$\eta=(n+1)/(0.19\times n)$.
For red supergiants (convective stars) $1 \leq n \leq 3/2$ leading to $8.7\leq \eta \leq
10$, 
while for blue supergiants and
Wolf-Rayet (radiative) stars $n=3$ leading to $\eta=7$. Unfortunately, our data
lack the precision to discriminate between these two possibilities.
Since theoretically $m\le 1$ we conclude that the mean velocity of the shock
(computed as $R/t$ where $t$ is the time since explosion) is 
about 21,000 km s$^{-1}$.

Next we can place a lower limit on the total energy. In
principle, in addition to the energy in the electrons and in the
magnetic field some energy can be carried by protons. 
We cannot
estimate the proton energy. Thus the lowest estimate of the energy is
given
by the equipartition analysis and excluding that carried by protons (ions): 
\begin{equation} 
E_{\rm min}=\frac{B_{p}^{2}R_{p}^{3}}{6}\left(1+f_{\rm eB}\right)f
\end{equation}
This yields (assuming $f_{\rm eB}=1$ and $f=0.5$) $E_{\rm min}\approx [1.6, 1.8, 2.9, 3.8, 5.8]\times 10^{45}$
erg at epochs of June 4, 5, 7, 9, and 11, respectively.

\subsection{Time-dependent solution}

We now fit the data to a comprehensive multi-epoch SSA model (as
described by Equations~\ref{eq:S} and \ref{eq:tau}). The resulting fit parameters are
$\alpha=-1.15$, $\beta=-0.96$, $K1=453.43$, $K5=1.9772\times 10^{5}$, and
$\delta^{''}=-3.42$. This implies an electron power law distribution with 
$p\approx3$ and $R\propto t^{0.94}$ which is consistent with what we found in the single epoch
analysis we performed above.
\begin{figure}
\centering
 \includegraphics[width=0.5\textwidth]{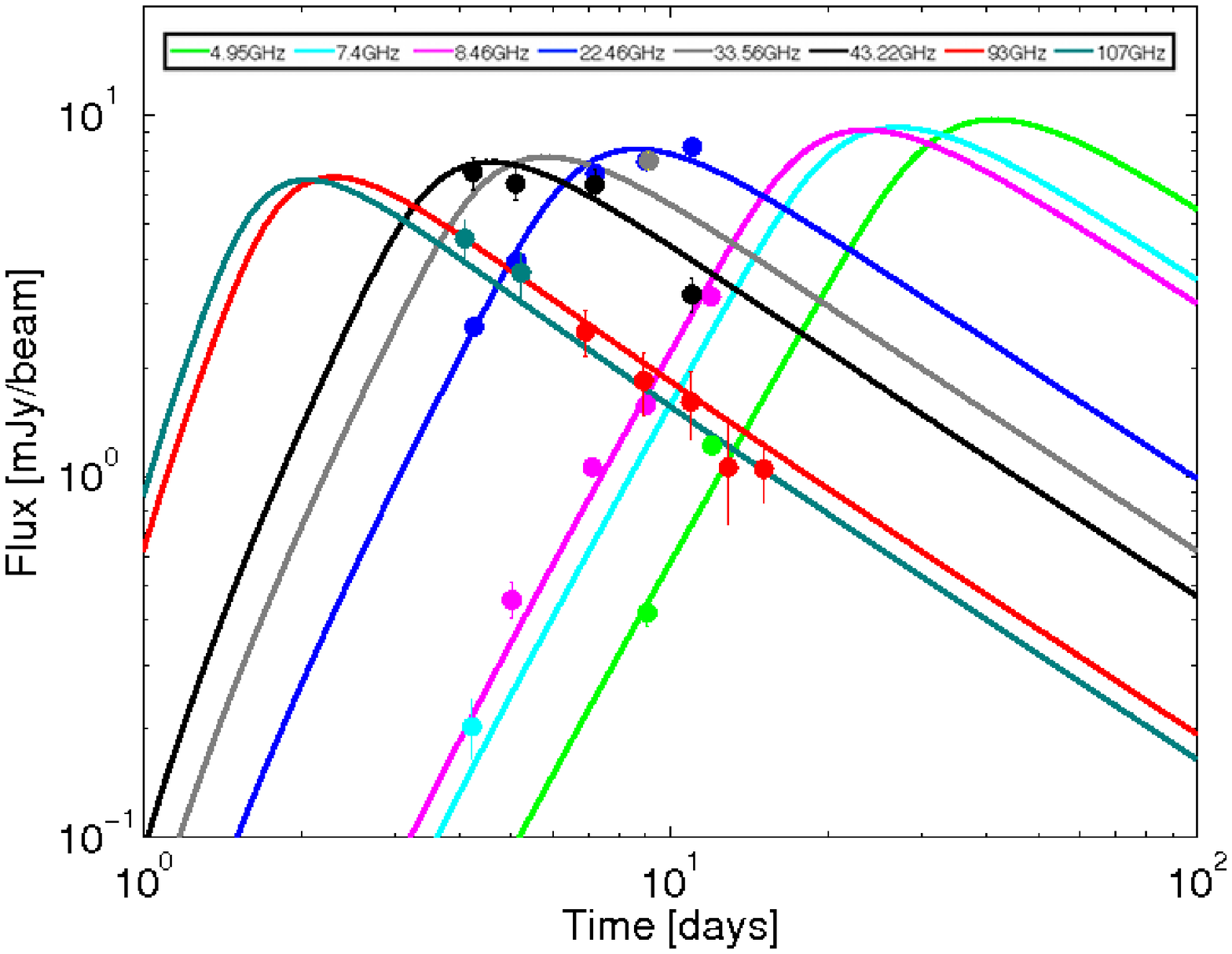}
 \includegraphics[width=0.5\textwidth]{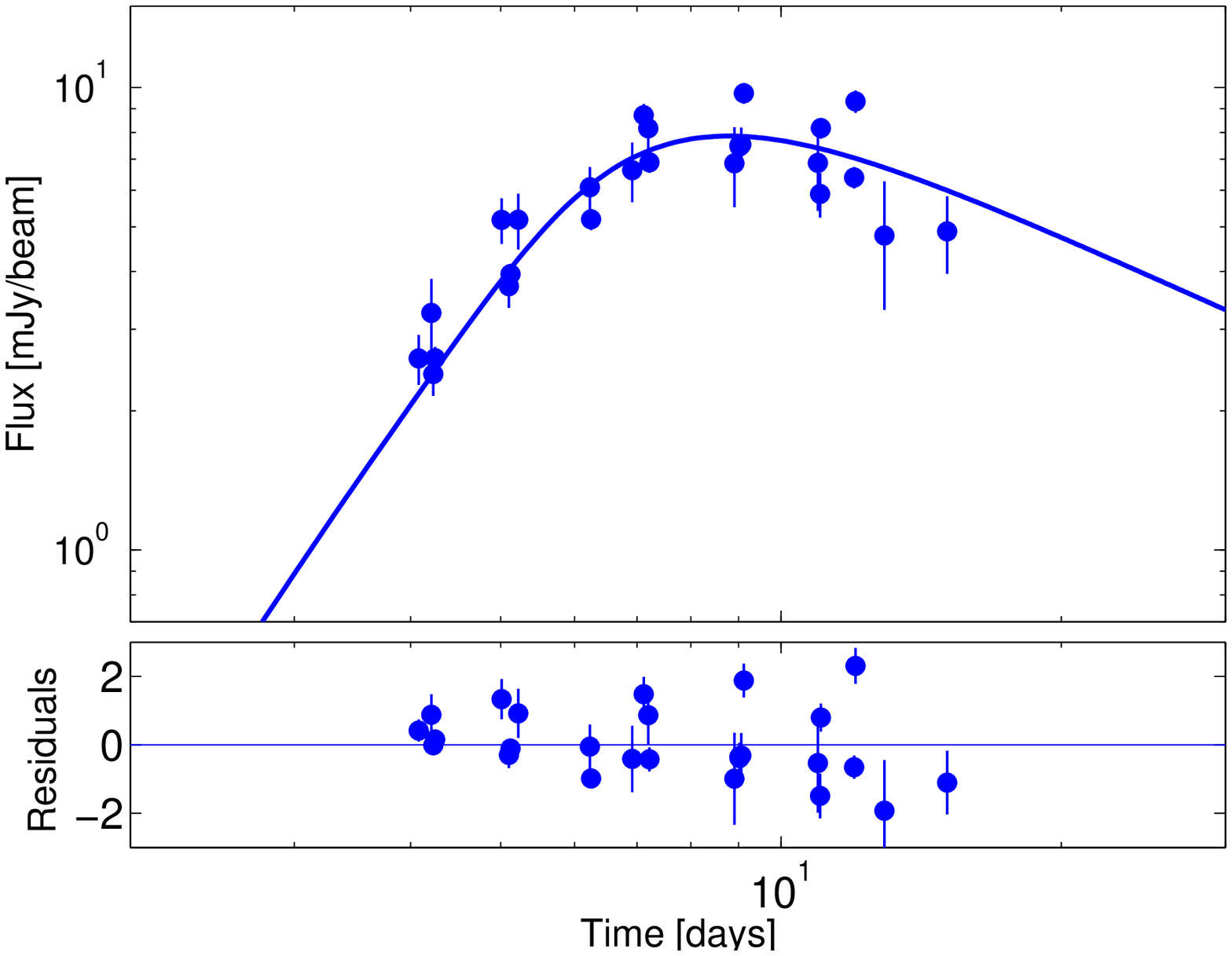}

\caption{Radio emission as a function of time. In the upper panel the
  result of the time-dependent fit (solid curves) is presented separately for each
  of the following frequencies
  $\nu=[107, 93, 43.22, 33.56, 22.46, 8.46, 7.4, 4.95]$ GHz from left to
  right, respectively. In the lower panel the time-dependent fit and
  its residuals for a frequency of $\nu=22.46$ GHz is shown. The measurements
  presented in the lower panel were shifted to the latter frequency,
  assuming the fit parameters from \S 3.2.}
\end{figure}

Assuming the CSM was formed by a constant wind mass-loss\footnote{We assume that the circumstellar matter is ionized. This
  is probably not an issue for blue super giants or Wolf-Rayet
  progenitors. For red super giants, the strong UV flash should
  provide some amount of ionization (see Chevalier et al. 1982).
}, 
the CSM
density structure will have the form $\rho_{\rm CSM}\propto r^{-2}$. We next calculate the
electron number density which is given by
$n_{e}=f_{\rm eB} \frac{(p-2)}{(p-1)}\frac{B^{2}}{8 \pi \gamma _{m} m_{e}
  c^{2}}$, where $\gamma _{m}$ is the minimum electron lorentz
factor (Soderberg et al. 2006). The resulting electron density is $n_{e}\approx 1.6-2.5\times
10^{5} \left(\frac{R}{10^{15}~{\rm cm}}\right)^{-2}{\rm cm}^{-3}$. In order to estimate the mass loss rate of the
progenitor via wind prior to the explosion, $\dot{{\rm M}} = 4\pi
R^{2} n_{e} m_{p} v_{w}$, an assumption about the wind velocity,
$v_{w}$, has to be made. Following
Chevalier \& Fransson (2006), instead of assuming a wind velocity we
will scale the mass loss rate by it and define a new parameter, 
$A={\dot M}/{4\pi v_{w}}$. In the case of SN\,2011dh, $A\approx
5\times 10^{11}$ g\,cm$^{-1}$, which is a factor of $3$ smaller than the value
derived by Soderberg et al. (2011) and Krauss et al. (2012). 
This value of $A$ corresponds to a mass-loss rate of $10^{-7}\times
(v_{w}/10 ~{\rm km~s^{-1}})$ M$_{\odot}$ yr$^{-1}$, 
consistent with a broad range of
massive progenitor stars, from red supergiants to compact Wolf-Rayet
stars.


\section{X-ray Observations}
\label{sec:Xray}

\begin{figure}
\plotone{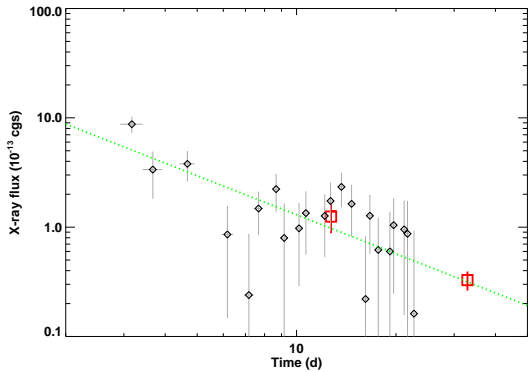}
\caption[]{\small \xray\ light-curve of \sn\ (0.3--10 keV) as determined from
  \swift\ XRT observations (black diamonds) and the high-level analysis software at the
  UK Swift Science Data Centre (Evans et al. 2007). Error bars are $1\sigma$. Data have been binned
  to half-day intervals, with bins having less than 5\% coverage
  dropped from analysis, and converted from counts to flux using the
  results of our \xray\ spectral fits. The two red squares are \chandra
  ~measurements converted to (0.3--10 keV) band. A fit to the temporal decay
  is indicated as a single power-law decay (dotted green
  line). Counts-to-flux conversion factors (\swift ~observations) for
  the two spectral epochs (the first 3 bins and the rest of the
  bins) are $c_{1,a} =
8.3\times 10^{-11}$ \cgsflux\ \perct\ and $c_{2,a} = 5.3\times 10^{-11}$ \cgsflux\ \perct, respectively, with respect to 0.3--10\,keV
absorbed flux, and $c_1 = 8.8\times 10^{-11}$ \cgsflux\ \perct\ and
$c_2 = 6.7\times 10^{-11}$ \cgsflux\ \perct, respectively, for
0.3--10\,keV unabsorbed flux.  The latter are the counts-to-flux
conversion factors we use to generate the \xray\ flux light-curve for
\sn.}
\label{fig:snxray}
\end{figure}

\sn\ was observed with the {\it Swift} X-ray telescope (XRT; Burrows et al. 2003) in a
series of observations beginning on 3.5 June 2011.  We
analyzed all observations
taken through 23~June 2011 using the pipeline software of the UK Swift
Science Data Centre at the University of Leicester\footnote{UK Swift
  Science Data Centre: \url{http://www.swift.ac.uk/}}, UK SSDC
hereafter (Evans et al. 2007). To generate the \xray\ light curve (see
Figure~\ref{fig:snxray}), we fix the source position at the
known location of the SN and extract a light curve with half-day
binning (43.2~ksec per time bin) using an aperture of
35.4\arcsec. For background subtraction, we determine the time-average count rate in this identical aperture, $1.29\pm 0.14$
cts \perksec\ (0.3--10 keV), from pre-\sn\ \swift\ data.

The SN \xray\ emission appears to fade rapidly at early times, so we
split the observations in two for purposes of \xray\ spectral
analysis, fitting data from the first three time bins separately from
the rest. The data are then fit to
\xray\ spectral models in XSPEC v 12.7.0 (Arnaud 1996), using a common Hydrogen
column density ($\nh$) across the full duration of the \swift\ observations and
allowing the power-law photon ($\Gamma$) index and time-average flux to vary
between the two data-sets. The resulting fit yields the following
spectral parameters, where uncertainties here and below are quoted at
90\%-confidence: $\nh = 1.1^{+1.3}_{-1.1}\times
10^{21}$\,\percmsq; $\Gamma_1 = 1.1\pm
0.4$; and $\Gamma_2 = 1.9^{+0.3}_{-0.4}$. The
results suggests a softening of the \xray\ emission over the course
of the \swift\ observations but note that the photon index overlaps for $\Gamma=1.5$ at the 90\%-confidence intervals.

Prior to analyzing the resulting \xray\ flux light-curve
(Fig.~\ref{fig:snxray}), we incorporate \chandra\ \xray\ flux
measurements from Pooley (2011) and Soderberg et al. (2011); these are
converted to our chosen 0.3--10\,keV bandpass using the authors'
chosen spectral models and the WebPIMMS tool\footnote{WebPIMMS tool:
  \url{http://heasarc.gsfc.nasa.gov/Tools/w3pimms.html}}.
Analyzing the resulting \xray\ light-curve, we find that
it may be adequately characterized (in a $\chi^2$ sense) as a
monotonic power-law decay emission. The single power-law
temporal decay index is $\alpha_{X} = -1.2\pm 0.2$ over the interval
  of \swift\ and \chandra\ observations. The \chandra\ data confirm the
  absence of significant contaminating source emission within the
  \swift\ XRT aperture (Pooley et al. 2011; Soderberg et al. 2011).


\section{Combined Radio-X-ray-optical analysis}
\label{sec:Combined}

The X-ray photon index is 1.1 (early times) and 1.9 (late times). The high
frequency radio spectral index is $-1$ which corresponds to a photon index
of $2$. The X-ray emission is elevated by a factor of 50 relative to extrapolation
of the SSA spectrum.  Thus we conclude that the X-ray emission is not a result
of synchrotron emission. Moreover, as already shown by Soderberg et
al. (2012), thermal free-free X-ray emission is ruled out. Another
explanation for the X-ray emission is that it arises
from inverse Compton (IC) scattering of optical photons.\footnote{
The SSA model yields $B$ of a few Gauss. A $\nu_p$ of 30 GHz would
require a Lorentz factor, $\gamma\sim 33$. These electrons
can inverse scatter SN optical photons to the X-ray band.}

Assuming that the X-ray emission arises from IC, the SSA and IC formulation of Katz (2011) can be used to infer
$R$ and $B$. For $p=3$
\begin{equation}
R\propto D
\left(r_{\rm SIC}F_{t}\right)^{1/10}\left(S_{\nu ,abs}\right)^{2/5}\left(\nu\right)^{-1}
\end{equation}
\begin{eqnarray}
B &=& 0.48 
\left(\frac{r_{\rm SIC}F_{t}}{10^{-12}{\rm erg~cm}^{-2}{\rm
      sec}^{-1}}\right)^{2/5}\left(\frac{S_{\nu ,abs}}{{\rm
      mJy}}\right)^{-2/5}\nonumber\\
& &\left(\frac{\nu}{10~{\rm GHz}}\right)
~{\rm G},
\end{eqnarray}
where $r_{\rm SIC}=\nu S_{\nu,syn}/(\nu S_{\nu, IC})$,  $S_{\nu,
  IC}$ is the IC flux, $S_{\nu, syn}$  is the synchrotron flux,
$S_{\nu, abs}$ is the SSA flux, and $F_{t}$ is the optical flux.

Applying the Katz (2011) equations to the optical (Arcavi et al. 2011b),
X-ray and radio measurements on June 7, 9 and 11 yields
$B\approx [0.77, 0.68, 0.48]\,$G. The estimated error (obtained
from application of the rule of error propagations) is  
$\approx 15\%$. These values of $B$ are
$\approx 4.2$ smaller than the corresponding estimates
obtained from an equipartition analysis (\S\ref{sec:equip}).
Nominally, by using equation~\ref{eq:B} again, $f_{\rm eB}\approx 500$ to 1700 with $10^3$ as
a reasonable mean (which we adopt). However, as can
be seen from Equation~\ref{eq:R} this only results in a 30\% decrease
in the value of $R$, relative to that obtained from equipartition analysis
(\S\ref{sec:equip}). Thus, the mean velocity is $R/t\approx 1.5\times 10^9\,$cm\,s$^{-1}$.
The electron density is now higher by a factor of $\sim
50$ and therefore the mass-loss rate is higher by a factor of
$\sim 20$ (for a fixed wind velocity).

\section{Conclusions \&\ Ramifications}
\label{sec:Conclusions}

In this paper we present the earliest millimeter- and centimeter-wave monitoring
observations of the Type IIb supernova SN\,2011dh in the galaxy M51,  
between four days and $12$ days after the event. Using the wide
frequency coverage of the EVLA 
(4 to 43 GHz) and CARMA (100 GHz) we measure the key parameters
of the synchrotron self-absorbed spectrum. The radio observations were accompanied
by extensive X-ray observations by {\it Swift}. The 
X-ray emission, argued to be inverse Compton scattering of the SN optical
photons by the relativistic electrons that produced the radio emission, 
 combined with the radio observations,  allow us to relax the equipartition
assumption and track the radius and circumstellar density.

We infer a mean velocity of $R/t \approx 1.5\times
10^9\,$cm\,s$^{-1}$, for the SN shockwave, with an uncertainty entirely dominated by the limitations
of the theoretical framework. Within the framework of the model the velocity is
1.32 to 1.68 $\times 10^9\,$cm\,s$^{-1}$.  This value range is  larger
than the $\sim 10^9\,$cm\,s$^{-1}$
expected for an extended progenitor (red supergiant) but smaller than
the $\sim 3\times 10^9\,$cm\,s$^{-1}$ 
expected for a compact progenitor. 
This may have important implications for the evolution leading to the formation of the progenitors
of SNe IIb: if SNe IIb are split into two distinct, well-separated classes, that would suggest they may
arise from two different evolutionary scenarios. On the other hand, if additional intermediate objects are found
, that may suggest a continuum of objects between eIIb and cIIb, a single
progenitor class may be favored.

We use the joint SSA+IC model analysis to infer the evolution of radius
with time, $R\propto t^m$. We find $m=1.14\pm 0.24$.  The error bars are
too large to directly constrain the physical nature of the envelope of the exploding
star (radiative versus convective).  Assuming no additional systematics the 
combination
of the early measurements with observations over several  
months may provide the means to determine the deceleration
phase of the shock-wave in a more precise way and indirectly may shed more light
on the density profile of the outer layers of the progenitor.

Soderberg et al. (2012) infer a much higher shock-wave velocity, $3\times
10^9$\,cm\,s$^{-1}$, when assuming equipartition. As noted in \S1
their analysis primarily rested on the X-ray data and optical data\footnote{
The analysis uses the method of Chevalier
\& Fransson (2006) which describes the IC emission as a function of optical
luminosity, $f_{\rm eB}$, shock-wave velocity, mass-loss rate, wind
velocity, and time.} 
(since the
radio data was quite sparse, day 4 and day 17). The limitations 
naturally propagate to the robustness of their inferences. This caveat notwithstanding, 
these authors find $f_{\rm eB}\equiv 30$, a factor of $\sim 30$ lower
than the value we find, suggesting a shockwave velocity of $2.5 \times 10^9$\,cm\,s$^{-1}$.  At least within the framework of the SSA+IC
 model we believe that our inference is more robust, thanks to our comprehensive
 radio and X-ray data sets. Moreover, as mentioned above, there is a
 factor of $3$ difference between the derived equipartition mass-loss
 rate value in our analysis compared to Soderberg et al. (2012) The origin of
 this difference is in the way this value is calculated and in the
 assumptions made. While Soderberg et al. (2012) assume $\epsilon_{e}=0.1$
 we do not make such an assumption. Our measurements suggest
 $\epsilon \sim 0.3$ which accounts for the different mass-loss rate
 values. 
 
Another set of comprehensive radio data was presented by Krauss et
al. (2012) for days $>20$ after explosion. In their analysis they
assumed equipartition only and found an average velocity of $\approx
25000$ km/s. Furthermore, assuming $\epsilon_{e}=0.1$, they found a
mass-loss rate $3.5$ times greater than what we find. In view of these
differences we analysed the data presented in Krauss et al. (2012) in
the same manner that we analysed the early data set we present above. Our
analysis suggests an average shockwave velocity of $18,500\pm 2000$
km/s at later times. Our derived lower velocity can be explained by a
few factors. First, our fitting method allows the electron energy
power-law index, $p$, to vary. We find an average value of $p=2.8$,
while Krauss et al. (2012) keep the power-law index constant at $p=3$. This
leads to slightly different values of the peak flux and frequency
which in turn leads to a lower value of the velocity. Another factor
that contributes to the difference between Krauss et al. (2012) and our
velocity value is the different coefficient used in Equation
$3$. While we are using the coefficient from Chevalier et al. (1998),
Krauss et al. (2012) use a coefficient which is larger by $20\%$. When
using Krauss et al. (2012) values for the peak frequencies and fluxes in
Equation $3$ using our coefficient, we find an average shockwave
velocity of $21,000$ km/s which is consistent with our results
given our errors.

In light of the above uncertainty in the
radius coefficient and in the fitting method
(constant vs. varying electron power-law index $p$), we note that the uncertainty
in derived values of the shockwave properties is greater than the
error measurements alone. The different uncertainties, combined, can
be as high as $40-50 \%$. Therefore, any conclusion based on the
absolute derived value of the shockwave properties such as its
velocity, is weakened. The time evolution of these properties is less
sensitive to the above uncertainties and therefore may provide a more
robust diagnostic.

On another matter, thanks to broad banding, 
improved receivers, and a flexible correlator, the Expanded VLA (EVLA) has
sensitivity gains ranging from factor of 2 to 10 in the 1--40 GHz band. 
Separately, there have been relentless and steady improvements in the
continuum sensitivity of millimeter wave arrays (CARMA, PdBI) with the
ALMA now offering an order of magnitude increase in sensitivity in the 
millimeter and sub-millimeter bands. 

These great improvements offer powerful diagnostics in two ways. First, 
both synchrotron self-absorption and free-free absorption (from the circumstellar
medium) depends strongly on frequency. 
High frequency observations can be sensibly undertaken at very early times.
(EVLA bands are usually self-absorbed at, say, day 1). Thus, early millimeter
wave observations can probe the fastest moving ejecta (since in a homologous
flow the fastest moving ejecta is at the greatest radius). Next, the intrinsic
SSA spectrum is modified by the external free-free optical absorption. This is
best measured by comparing lower frequency measurements to higher frequency
measurements. A clear detection of free-free absorption gives a model independent
measure of the density of the circumstellar matter. 

These two diagnostics motivated our CARMA+EVLA effort. 
The toy model given in Appendix A shows that a future
SN, such as SN\,2011dh, if observed at, say, day 1, would allow us to meet at least one of
these two goals. Given that surveys such as PTF are now moving to even
faster cadence we would expect these two goals to be realized in the very
near future.

\section*{Acknowledgments}

We thank the CARMA and EVLA staff for promptly scheduling this target
of opportunity. The National Radio Astronomy Observatory is a facility
of the National Science Foundation operated under cooperative
agreement by Associated Universities, Inc. This work made use of data supplied by the UK Swift
Science Data Centre at the University of Leicester. PTF is a fully-automated, wide-field survey aimed at a systematic exploration of explosions and variable phenomena in optical wavelengths. The participating institutions are
Caltech, Columbia University, Weizmann Institute of Science, Lawrence
Berkeley Laboratory, Oxford and University of California at
Berkeley. The program is centered on a 12Kx8K, 7.8 square degree CCD
array (CFH12K) re-engineered for the 1.2-m Oschin Telescope at the
Palomar Observatory by Caltech Optical Observatories. Photometric
follow-up is undertaken by the automated Palomar 1.5-m
telescope. Research at Caltech is supported by grants from NSF and
NASA. The Weizmann PTF partnership is supported in part by the Israeli
Science Foundation via grants to A.G. Weizmann-Caltech collaboration
is supported by a grant from the BSF to A.G. and S.R.K. A.G. further
acknowledges the Lord Sieff of Brimpton Foundation. CJS is supported
by the NASA Wisconsin Space Grant Consortium.
FEB acknowledges support from CONICYT-Chile under grants FONDECYT
1101024 and FONDAP-CATA 15010003, Programa de Financiamiento Basal,
the Iniciativa Cientifica Milenio through the Millennium Center for
Supernova Science grant P10-064-F, and Chandra X-ray Center grants SAO
GO9-0086D and GO0-11095A. MMK acknowledges support from the Hubble Fellowship and the
Carnegie-Princeton Fellowship. NP acknowledges partial support by
STScI-DDRF grant D0001.82435. Research at the Naval Research
Laboratory is supported by funcding from the Office of Naval
Research. S.B.C.~acknowledges generous financial assistance from Gary
\& Cynthia Bengier, the Richard \& Rhoda Goldman Fund, the Sylvia \&
Jim Katzman Foundation, the Christopher R. Redlich Fund, the TABASGO
Foundation, NSF grants AST-0908886 and AST-1211916. We thank the anonymous referee for his constructive
comments. 

\appendix
\section{Why early radio (cm and mm) observations of SNe are important}

The free-free optical depth is
\begin{equation}
	\tau_{\rm ff}=3.3\times 10^{-7}T_4^{-1.35}\nu_{\rm GHz}^{-2.1} {\rm EM}
\end{equation}
where $T=10^4T_4$ is the electron temperature (in degrees
Kelvin)\footnote{We use the convention of $X_n=X/10^n$ where it is
assumed, unless explicitly specified, that the units are CGS.},
$\nu_{\rm GHz}$ is the frequency in GHz and EM is the emission
measure, the integral of $n_e^2$ along the line of sight, and in
units of cm$^{-6}$\,pc.

For a star which has been losing matter at a constant rate, $ \dot M$,
the circumstellar density has the following radial profile:
\begin{equation}
\rho(r)=n(r)\mu = \frac{\dot M}{4\pi r^{2} v_{w}}
\end{equation}
where $v_{w}$ is the wind velocity, $r$ is the radial distance from
the star, $n$ is the particle density and $\mu$ is the mean atomic
weight of the circumstellar matter. Thus $n(r) \propto r^{-2}$. We assume that the circumstellar matter is ionized. This is probably not an issue for blue super giants or Wolf-Rayet progenitors. The strong UV flash for red super giants provide some amount of ionization and so a specific check needs to be done for such progenitors.
 The emission measure from a radius, say,
$r_1$ to infinity is
	\begin{eqnarray}
	{\rm EM} &=& \int_{r_1}^{\infty} n_*^2\Big( \frac{r}{r_*}\Big)^{-4}dr
	=\frac{1}{3}n_*^2r_*\Big( \frac{r_1}{r_*}\Big)^{-3}
	\end{eqnarray}
where $n_*$ is the density of electrons at radius $r_*$.  
We choose the following normalization for these two quantities: $n =
10^6 n_6$ and $r = 10^{15} r_{15}$. The corresponding expression for the
emission measure is 
\begin{equation}
{\rm EM}=1.08 \times 10^8 n_{*6}^{2}r_{*15}\Big( \frac{r_{15}}{r_{*15}}\Big)^{-3}
\end{equation}

Using the above equation for EM and substituting $r_{15}=0.0864 \times v_9 t_d$ the free-free optical depth is
\begin{equation}
\tau_{\rm ff}=5.5\times
10^{4}n_{*6}^{2}r_{*15}^{4}v_9^{-3}t_d^{-3}T_4^{-1.35}\nu_{\rm
  GHz}^{-2.1} 
\end{equation}

We use SN\,2011dh as an example and set $v_9=1.5$, $r_{*15}=1$, and
$n_{*6}=0.2$. In this case the free-free optical depth is $\tau_{\rm
  ff}=6.5\times 10^{2} T_4^{-1.35}\nu_{\rm
  GHz}^{-2.1}$. Setting the LHS to unity and assuming $T_4=1$ yield the run of the
frequency at which {\it qualitatively} the optical depth is unity:
\begin{equation}
\nu_{\rm GHz}({\rm ff})=21.8 t_d^{-1.43}
\end{equation}

The
SSA optical depth is described by
\begin{equation}
\tau_{SSA}=K5 \left(\frac{\nu}{\rm 5\,GHz}\right)^{\alpha-2.5}
\left(\frac{t-t_{0}}
{\rm 1\,day}\right)^{\delta^{''}},
\end{equation}
where both $K1$ and $K5$ are proportionality constants that can be
determined by fitting the data, and $\delta^{''}$ describes the time
dependence of the optical depth.
In \S 3.2 we found the following parameters for SN\,2011dh: $\alpha=-1.15$,$K5=1.98\times 10^{5}$, and
$\delta^{''}=-3.42$. The SSA optical depth in this case is 
\begin{equation}
\tau_{\rm SSA}=7 \times 10^{7} \nu_{\rm GHz}^{-3.65} t_d^{-3.42}
\end{equation}
Setting $\tau_{\rm SSA}$ to unity yields the run of the peak SSA
frequency as a function of time
\begin{equation}
\nu_{\rm GHz}({\rm SSA})=141t_d^{-0.937}
\end{equation}

Armed with equations A5 and A8 with the fitted parameters we
found in \S 3.2, we can now look into the value of even earlier
mm-observations (see Figure 5). 
\begin{figure}[!ht]
\centering
\includegraphics[width=0.75\textwidth]{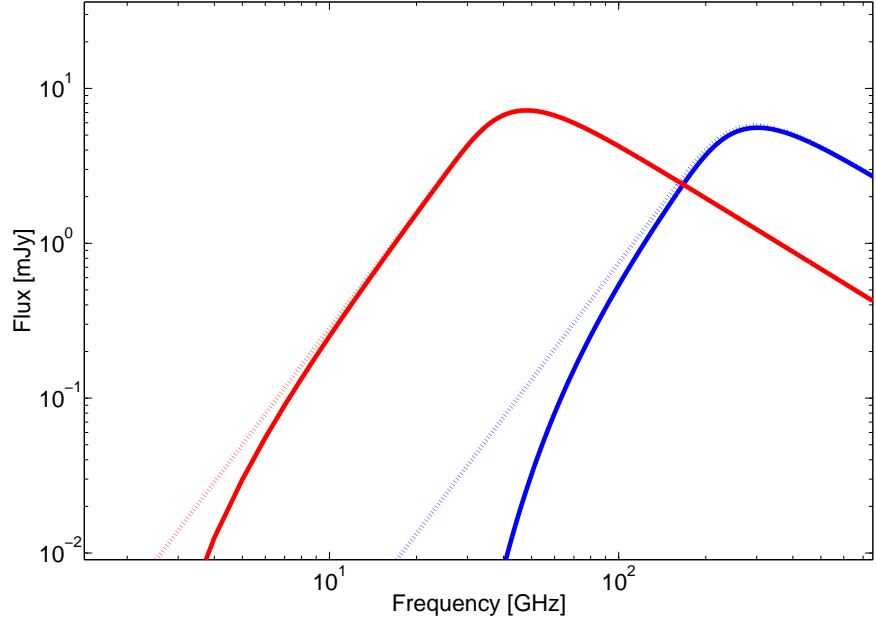}
\caption{Theoretical radio spectra of SNe including synchrotron-self absorption and with (solid lines) or without
(dashed lines) free-free absorption. Blue and red lines are for days 1
and 4 after explosion, respectively. The above spectra were produced
using Equations 5, A5 and A8 with the fitted parameters found in $\S
3.2$ for SN\,2011dh.}
\end{figure}
If had we observed SN\,2011dh on day 1 the EVLA observations would have shown clear
signature for free-free absorption. We also note that $\tau_{\rm ff}$
is proportional to $n_*^2r_*$ whereas the SSA optical depth is a
different function of $n_*$ and $r_*$. Thus we can, in principle, obtain a different measure of $n_*$ and $r_*$.




{}

\end{document}